\documentclass[aps,prl,superscriptaddress,showpacs,twocolumn]{revtex4}

\usepackage[dvips]{graphicx}

\begin{document}

\bibliographystyle{apsrev}

\title{Dynamic density functional study of a driven colloidal particle in polymer solutions}

\author{F. Penna}
\affiliation{%
Departamento de F{\'{\i}}sica Te\'orica de la Materia
Condensada,
Universidad Aut\'onoma de Madrid, E-28049 Madrid, Spain}
\author{J. Dzubiella}
\affiliation{%
University Chemical Laboratory,\\
Lensfield Road,Cambridge CB2  1EW, United Kingdom
}
\author{P. Tarazona}
\affiliation{%
Departamento de F{\'{\i}}sica Te\'orica de la Materia
Condensada,
Universidad Aut\'onoma de Madrid, E-28049 Madrid, Spain}

\date{\today}

\begin{abstract}
The Dynamic Density Functional (DDF) theory and  standard 
Brownian dynamics simulations (BDS)
are used to study the drifting effects of a colloidal particle in a polymer solution, 
both for ideal and interacting polymers. 
The structure of the stationary density distributions 
and the total induced current are analyzed for different drifting rates.
We find good agreement with the BDS, which gives support to the
assumptions of the DDF theory. The qualitative aspect of the density
distribution are discussed and compared to recent results for driven
colloids in one-dimensional channels and to analytical expansions for the 
ideal solution limit.
\end{abstract}


\maketitle

\section{Introduction}

Mixtures of colloids and non-adsorbing polymer coils have attracted much attention over
the past decade both experimentally and theoretically \cite{poon}. They
provide an excellent model system to understand the generic equilibrium
and nonequilibrium physics of multicomponent colloidal mixtures due to
the fact that the interactions between the constituents can be
tailored \cite{pusey}. The
theoretical approaches to the study of the equilibrium phase behavior
of colloid-polymer mixtures were largely based on the Asakura-Oosawa
(AO) \cite{AO} model, in which the chains are modeled as ideal
particles experiencing a hard-core repulsion with the
colloids. Recently, extensive  computer simulations \cite{Ard} revealed
qualitative differences in the phase diagrams when interactions
between the polymers are included. However, the off-equilibrium behavior
of these systems is far from being understood. In this paper, we
present results based on a recently proposed dynamical density
functional (DDF) formalism \cite{OurJCP,OurJPCM} and we demonstrate that the latter is
capable of describing out-of-equilibrium diffusive processes at the
Brownian timescale.  The  advantage of the DDF theory is the fact
that particle interactions are included once a good approximation for
the equilibrium functional is known and it is well suited to
treat different external potentials.  
 
The system under consideration is a  colloidal particle
being dragged at a constant rate $c$ (e.g. by gravitation,  electric or
magnetic fields, or by optical clamps \cite{forces})
through a solution of polymers in a light solvent. The spherical
colloid  is represented by an  external potential $V_{{\rm ext}}({\bf
    r},t)=V_{{\rm ext}}(|{\bf r'}|)$,
where ${\bf r'}\equiv {\bf r}-ct{\bf\hat{z}}$ is the  coordinates 
in the reference framework of the colloidal particle.
The solvent provides the rest-framework for the Langevin dynamics of the
polymers, which have a mobility $\Gamma_0$, connected through the Einstein relation
to the diffusion constant and the inverse thermal energy, $\beta=(k_{B}T)^{-1}$.
Assuming that the polymer gyration radius and the colloidal particle
have similar size, $\sigma \sim 10^{-7}$m, and with the viscosity of 
a typical solvent at room temperature, the natural units for the shifting rate 
$\Gamma_0/(\beta \sigma)$ would be in the range of $10^{-4}$m/s, and we may 
safely neglect the hydrodynamic effects of the light solvent.
The much heavier globular polymers
would feel the competing effects of the solvent, as rest framework for their
Brownian dynamics, and the moving colloidal particle drifts 
with respect to the polymers at a constant rate (drift velocity) $c$.
The deterministic Dynamic Density Functional (DDF) theory \cite{OurJCP,OurJPCM}, is an extension of the Density Functional (DF) formalism to off-equilibrium systems,which includes  exactly the ideal gas and the external potential
contributions to the free energy, and it 
represents the correlations out of 
equilibrium by those of an equilibrium system with the same 
density distribution. With this hypothesis, 
and the interpretation of the density $\rho({\bf r},t)$ 
as the average of the instantaneous density over 
the random noise in the molecular Langevin dynamics, the theory enables 
the use of the well developed approximations for the 
equilibrium  Helmholtz free energy 
in DF theory, which are usually split into the ideal gas and 
interaction contributions, 
${\cal F}[\rho]={\cal F}_{\rm ideal}[\rho]+\Delta{\cal F}[\rho]$.
 The central DDF equation for the time-dependent density distribution is
\begin{eqnarray}
\frac{\partial \rho ({\bf r},t)}{\partial t}= \Gamma_0
\ \nabla \left[ \rho({\bf r},t) \ \nabla
\left( \frac{\delta{ \cal F}[\rho]}
{\delta \rho({\bf r},t)} +V_{{\rm ext}}({\bf r}') \right) \right].
\label{DDF}
\end{eqnarray}

 From any initial distribution of polymers, the time evolution would
take the system towards to a stationary density distribution
$\rho({\bf r},t)= \rho({\bf r}- \hat{\bf z} c t)
\equiv \rho({\bf r'})$, shifting at the same rate $c$ as the external
potential. This distribution is the
most relevant property of the system and it corresponds to the solution
of the functional equation 
\begin{eqnarray}
\ \nabla \cdot \left[ \rho({\bf r'}) \ \nabla
\left( \frac{\delta{ \cal F}[{\rho}]}
{\delta \rho({\bf r'})} +V_{{\rm ext}}({\bf r'})+\frac{c z'}{ \Gamma_0
} \right) \right]=0,
\label{SDDF}
\end{eqnarray}
where the time dependence is fully adsorbed into the coordinate
${\bf r'}$. The solution of Eq.(\ref{SDDF}) has been explored for 
systems with one-dimensional (1D) dependence of the potential
barrier \cite{1d}, $V_{\rm ext}({\bf r'})=V_{\rm ext}(z')$, along the shifting 
direction; the similarities and the differences with the
Euler-Lagrange equation for the DF theory of equilibrium systems 
were analyzed there  both for ideal and interacting systems. However,
the three-dimensional (3D)  geometry of the present problem
requires a different analysis, since the zero-divergence requirement
for the brackets in Eq.(\ref{SDDF}) leaves open a much wider functional
space in 3D than in 1D.

\section{Non-interacting polymers}

Starting in the spirit of the AO model, for non-interacting polymers,
$\Delta{\cal F}=0$, Eq. (\ref{SDDF}) becomes a linear Fokker-Planck  
equation 
\begin{eqnarray}
\nabla^{2} \rho +
 \nabla(\rho \cdot \nabla\beta V_{k})=0 ,
\end{eqnarray}
 with a 
$"$kinetic potential$"$  given by  
$\beta V_{k}({\bf r'}) \, 
=\beta V_{{\rm ext}}({\bf r'})+\bar{c}z\,'$,
as a function of  the reduced shifting rate $\bar{c}\equiv \beta c/\Gamma_0$
with inverse length units.
In Fig. 1(a,b) we present numerical solutions for the density
distribution of the ideal case
with bulk density $\rho_0 \sigma^{3}=1$,
under the effects of the external potential
\begin{eqnarray}
V_{{\rm ext}}({\bf r'})=V_0 \ \exp( - |{\bf r'}/\sigma|^6),
\end{eqnarray}
with $\beta V_0=10$, to represent the soft repulsion between the polymers and
the shifting colloidal particle. The bulk density $\rho_{0}$ is the
value of the density far away from the external potential and is
obviously the same both for the equilibrium and the nonequilibrium, driven
system; moreover, for the ideal non-interacting system $\rho_0$ just
provides an arbitrary factor to $ \rho({\bf r'}) $.   

The density distribution around the external potential 
$\rho({\bf r'})$ has axial symmetry 
and exhibits a cap-like structure
with  $\rho({\bf r'}) >\rho_0$ in the {\it front}  ($z'\gtrsim \sigma$), formed by the polymers being pushed 
by the moving repulsive external potential. These particles escape around
the colloidal particle creating a skirt for $z'\lesssim -\sigma$, which,
together with the hole ( $\rho({\bf r'}) <\rho_0$) left behind 
by the potential, form a {\it wake} structure extending much further than
the {\it front} structure for $z'\gtrsim \sigma$. An increased shifting rate
( $\bar{c} \sigma=10$ in Fig.1b compared to $\bar{c} \sigma=1$ in Fig.1a), 
enhances these characteristics, with a 
higher density {\it front} 
and a narrower skirt reaching further away behind the shifting 
colloidal particle. 

The transverse integral of $\rho({\bf r}')-\rho_o$ over $x'y'$ plane
as function of $z'$ gives an excess of polymers 
in the {\it front} region
but it vanishes in the {\it wake} region, as the positive wings exactly
compensate the depletion close to the $z'$-axis. This is reminiscent of
the 1D result \cite{1d} for a shifting repulsive barrier with a
{\it front} density $\rho(z')=\rho_0+A \exp(- \bar c z')$
and no  {\it wake} , $\rho(z')=\rho_0$, behind. The absence of excess 
molecules in the {\it wake} seems to be a generic characteristics of
stationary states under constant bulk boundary conditions
with purely relaxative dynamics of the ideal gas molecules.
Using  cylindrical coordinates ${\bf r}'=(R,\phi',z')$
we now explore analytically the asymptotic forms of both the {\it front}
and the {\it wake} 
regions, where $V_{{\rm ext}}(r')=0$
reduces 
Eq.(\ref{SDDF}) to:
\begin{eqnarray}
\frac{1}{R}\frac{\partial}{\partial R }\left(R \frac{\partial \rho({\bf r'})
}{\partial R }\right)+
\frac{\partial^{2} \rho({\bf r'})
}{\partial z'^{2} }+\bar{c}\frac{\partial \rho({\bf r'})
}{\partial z'}=0.
\label{3Deq}
\end{eqnarray}
Through a  Hankel transform, the solution for the 
cylindrically  symmetric $\rho({\bf r'})$ is,
\begin{eqnarray}
\rho(R,z')=\rho_{0}+\int_{0}^{\infty}d\alpha \,\alpha\,f(\alpha)\,
  J_{0}[\alpha\, R] \, e^{-\beta \, z'};
 \label{3Dsol}
\end{eqnarray}
where  $\beta_{\pm}=\frac{\bar{c}}{2}\pm\sqrt{(\frac{\bar{c}}{2})^{2}+\alpha^{2}} $ and
$ J_{0}$ is the zeroth order Bessel function.
Far away from the external potential, a fast convergence of the Hankel 
components $f(\alpha)$  is observed, and  the relevant features come from
their behavior for \begin{math}\alpha\ll\bar{c}\end{math}. Hence we may
use  \begin{math}
\beta_{+}\approx\bar{ c }+\frac{\alpha^{2}}{\bar{ c }}\end{math} and 
\begin{math} \beta_{-}\approx -\frac{\alpha^{2}}{\bar{ c }}\end{math}, 
for the {\it front}  and the  {\it wake}  regions respectively. The expansion of 
$f_{\pm}(\alpha)$ 
as an even  polynomial function for small $\alpha$ and the zero  {\it wake}  
requirement lead to 
$ f_{-}(\alpha)\approx A_{1}\alpha ^{2}+ A_{2} \alpha ^{4}+...$ 
in the  {\it wake }, while at the {\it front} we expect  
$f_{+}(\alpha)\approx B_{0}+ B_{1}\alpha ^{2}+ B_{2} \alpha ^{4}+... $
Thus 
\begin{eqnarray}
\rho(R,z')\approx \rho_{0} + e^{-\bar{c} z'}e^{-\frac{a^{2}}{4}}[B_{0}\frac{w^{2}}{2}
\nonumber
\\
\ +B_{1}\frac{ w^{4}}{8}\left(a^{2}-4\right)
+ B_{2}\frac{ w^{6}}{32} \left(a^{4}-16 a^{2}+32\right)+... ],
\label{3Dsol1}
\end{eqnarray}
for $z'\gg \sigma $, and
\begin{eqnarray}
\rho(R,z')\approx \rho_{0} + 
e^{-\frac{a^{2}}{4}} [A_{1}\frac{ w^{4}}{8}\left(a^{2}-4\right )
\nonumber
\\
\ + A_{2}\frac{ w^{6}}{32} \left(a^{4}-16 a^{2}+32\right)+... ];
\label{3Dsol3}
\end{eqnarray}
for $z'\ll -\sigma$;
where
\begin{math}
w=\sqrt{\bar{c}/\mid z' \mid} \,\,\end{math} 
and\begin{math}\ \, \,\,a=wR\end{math}.\\ 
Although the amplitudes of these contributions depend on the particular
external potential, the asymptotic decay forms are generic.
For a fixed $z'$, the structure in the  transverse plane is given by a 
gaussian times a polynomial function. In our case the positive values
of $A_1$ and $B_0$  lead to a maximum {\it front}  density  
at $R=0$, while the leading term at the {\it wake} gives a minimum at
$R=0$ and maximum at $R= 2 ^{3/2} \sqrt{ \mid z' \mid/\bar{c}}$, 
producing the cap-like structure with  parabolic shape shown in Fig. 1(a,b).
Besides the  lack of $A_0$,  the qualitative difference between the 
advancing {\it front}  and the  {\it wake}  comes from the  exponential decay 
$ \exp(-\bar{c} z')$ behavior of the {\it front} , which restricts the 
density structure to the neighborhood of the external potential, as in the 
$1D$ result, while for fixed $a$ the  {\it wake}  structure  decays as  
inverse powers of $z'$, with a  $1/z'^{2}$  leading term.
We have tested  these analytical predictions with the numerical 
solution of Eq.( \ref{SDDF}), and got quantitative agreement for
$z'>1.25 \sigma$ ($\bar{c}\sigma=10$) and $z'>3 \sigma$ ($\bar{c}\sigma=1$) 
in the {\it front} region, and $z'<-7 \sigma$ in the  {\it wake}  region for any $\bar{c}$.
The contributions from higher order terms in Eq.(\ref{3Dsol3})
appear to be more important than in  Eq.(\ref{3Dsol1}), although 
the qualitative aspect of the  {\it wake}  is already well represented by the
first term in  Eq.(\ref{3Dsol3}).  

Nevertheless, we have to point that, contrary to the 1D case \cite{1d},
the {\it front} and the {\it wake} regions in our 3D system are in fact
connected through the regions, with small $|z'|$ but large $R$, were 
the external potential created by the shifting colloidal particle on the 
polymers vanishes; so that the solution of Eq.(\ref{3Deq}) should have a
{\it unique} analytic form, common to the {\it front}  and the  {\it wake}  region.
This has the obvious difficulty of using both the positive ($\beta_+$)
and the negative ($\beta_-$) decay constants for the positive ({\it front})
and negative ({\it wake}) values of $z'$, leading to exponential growth
of their respective contributions, which may only be canceled by the
appropriate behaviour of $f(\alpha)$ for large $\alpha$, beyond the
Taylor expansion used in  Eq.(\ref{3Dsol1},\ref{3Dsol3}). The good comparison of
our numerical solutions with the analytic results   Eq.(\ref{3Dsol1}) for the
{\it front} and  Eq.(\ref{3Dsol3}) for the {\it wake} reflects a local
asymptotic convergence which is quite useful to
understand the qualitative features of the stationary density distribution,
but which cannot be taken as an exact global asymptotic result.
It has to be pointed that the use of spherical, rather than cylindrical
coordinates to solve Eq.(\ref{3Dsol}) also leads to problems of convergence,
as the parabolic structure of the  {\it wake}  implies the entanglement of the
radial and the angular coordinates.
%%%%%%%%%%%%%%%%%%%%%%%%%%%%%%%%%%%%%%%%%%%%%%%%%%%%%%%  
\begin{figure}
\vspace{-1cm}
\includegraphics[width=80mm]{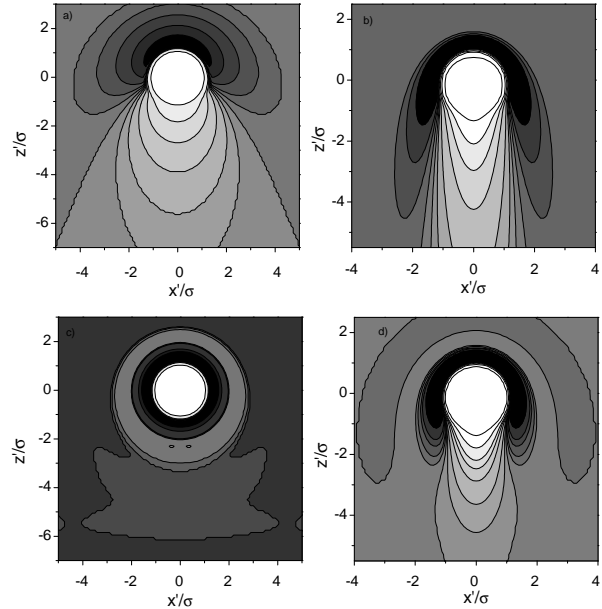}
\vspace{-2cm}
\caption{Steady state contour density field of ideal (a,b) and
  interacting  (c,d) polymers 
created by a driven colloidal particle. Shown
  is the density in the $x'z'$-plane where the center of the colloid
  is located. The colloid is  moved along the $z'$-axis at a velocity
  $\sigma\bar{c}=1$ (a,c) and $\sigma\bar{c}=10$ (b,d). Bright regions
  correspond to low densities while dark regions show high densities.}
\label{fig:1}
\end{figure}  
%%%%%%%%%%%%%%%%%%%%%%%%%%%%%%%%%%%%%%%%%%%%%%%%%%%%%%%  

\section{The effect of polymer interaction}   

For the case of interacting polymers the steric effects lead to
an effective repulsion between them, which we model by 
the ultra-soft gaussian pair potential \cite{gausspot,evanspre,evanscm,lang,Louis1,Louis2} 
\begin{eqnarray} \phi(r_{ij})=U_{0}\exp(-r_{ij}^{2}/\sigma^{2}),
\end{eqnarray}
where $r_{ij}$ is the interparticle distance and $\beta U_{0}=1$. Both
for equilibrium 
\cite{evanspre,evanscm,lang,Louis1} and dynamical properties \cite{joelikos1} the excess 
free energy density functional of this model has been successfully 
approximated by a pure {\it mean field} (or random-phase approximation
(RPA)) form
\begin{eqnarray}
\Delta {\cal F}[\rho]= {1\over 2} \int d^3{\bf r}  \int d^3{\bf r}'
 \phi(|{\bf r}-{\bf r}'|) \ \rho({\bf r}) \rho({\bf r}').
\label{mfa}
\end{eqnarray}
%%%%%%%%%%%%%%%%%%%%%%%%%%%%%%%%%%%%%%%%%%%%%%%%%%%%%%%%%%%%%%%%%%%
%%%%%%%%%%%%%%%%%%%%%%%%%%%%%%%%%%%%%%%%%%%%%%%%%%%%%%%%%%%%%%%%%%%

Instead of solving the integro-differential  Eq.(\ref{SDDF})
with this model, we have obtained the steady state distribution by the 
time integration of  Eq.(\ref{DDF}) from a uniform density initial state.
The stationary structure around the colloidal particle are reached with
short integration times, as presented in Fig. 1  for $\rho_0 \sigma^3=1$ (which represents a fairly dense polymer solution)
and  $\bar c=1\sigma$ (c) and $\bar c=10\sigma$ (d).  
At the low velocity $\bar{c} \sigma=1$ the influence of interactions
between the  polymers is very strong. The spherical layering
structure created around the colloidal particle by the polymer-polymer 
repulsion is much stronger than the {\it front} - {\it wake}  asymmetry induced by the
dragged colloid; the extension of the  {\it wake} 
behind the moving particle is strongly reduced by the much lower
bulk osmotic compressibility of the interacting system, which facilitates 
the filling of the axial hole by radial currents. At the higher
$\bar{c} \sigma=10$ shifting rate in Fig. 1(d) the effects of the interactions
are much weaker. Although there is still a clear shortening of the
 {\it wake} , explained by the lower osmotic compressibility, the main qualitative 
change with respect to the ideal solution result in Fig. 1(b) is that 
the layering created by the packing effects produces a double
cup-like structure, reaching further away from the $z'$-axis.

In Fig.2 we present a quantitative   view of our results; we
plot the polymer density as a function of the distance $R$ to the $z'$-axis for
fixed values of $z'/\sigma=\pm 1$.
The solution of the DDF approach is compared to standard Brownian
dynamics simulation (BDS) \cite{BDS}. Here, 
the stochastic Langevin equations for the overdamped colloidal motion
of $N$ particles with
trajectories  ${\bf r}_i(t)$ 
$(i=1,...,N)$ read as
\begin{equation}
\Gamma_{0}^{-1} \frac{{d{\bf r}_i}}{dt} = -{\bf \nabla}_{{\bf r}_i} \sum_{j\not= i} \phi
(\mid {\bf r}_i -{\bf r}_j \mid ) + {\bf F}_{\rm ext} (t) +
\Gamma_{0}^{-1}{\bf c} + {\bf
  F}_i^{(\rm R)}(t).
\label{langevin}
\end{equation}
There are different forces acting onto
the colloidal particles: first there is the force attributed to
inter-particle interactions, secondly there is the external field
${\bf F}_{\rm
  ext}$ due to the colloidal particle, $\Gamma_{0}^{-1}{\bf c}$ is the driving force, 
and finally  the random forces ${\bf F}_i^{(\rm R)}$ describe 
 the kicks of the solvent molecules acting onto the $i$th
 colloidal particle. These kicks
are Gaussian random numbers with zero mean, $\overline {{\bf
    F}_i^{(\rm R)}}=0$,
and variance 
\begin{equation}
\overline{({\bf F}_i^{(\rm R)})_{\alpha }(t)({\bf F}_j^{(\rm R)})_{\beta }(t')}={{2k_BT}{\Gamma_{0}}}
\delta_{\alpha\beta} \delta_{ij}\delta(t-t')
\label{variance}
\end{equation}
The subscripts $\alpha$ and $\beta$ stand for the three Cartesian components.
The simulations were carried out with $N=1000$ particles,  periodic boundary conditions in all
directions and box sizes $L_{x}=L_{y}=8\sigma$ and $L_{z}=16\sigma$.
After an equilibration time of $10^{5}$ timesteps, statistics were
gathered over a period of $2\cdot 10^{6}$ timesteps. 
In both, simulation and numerical solution of the DDF, the density is
averaged over rings at $z'$ and radius $R$ with a cross section
$\sigma^{2}/4$. 

The good agreement between the BDS data and the DDF results
gives support to both the mean field approximation (\ref{mfa}) used to 
describe the effect of the interactions, and to the DDF borrowing of 
the equilibrium intermolecular forces, as functionals of the instantaneous density
distribution \cite{OurJCP,OurJPCM}. The comparisons between the ideal solution
and the interacting system shows again that the
polymer interactions affect much more the features of
the density distribution for low $\bar{c}$ than for high $\bar{c}$.
For $\bar{c}\sigma=1$ the polymer layer around the colloidal
shows only a slight asymmetry for $\sigma \gtrsim z' \gtrsim -\sigma$,
while the kinetic effects in ideal gas create a maximum 
at the {\it front}  $z'\gtrsim \sigma$.
%%%%%%%%%%%%%%%%%%%%%%%%%%%%%%%%%%%%%%%%%%%%%%%%%%%%%%%%%%
\begin{figure}
\includegraphics[width=85mm]{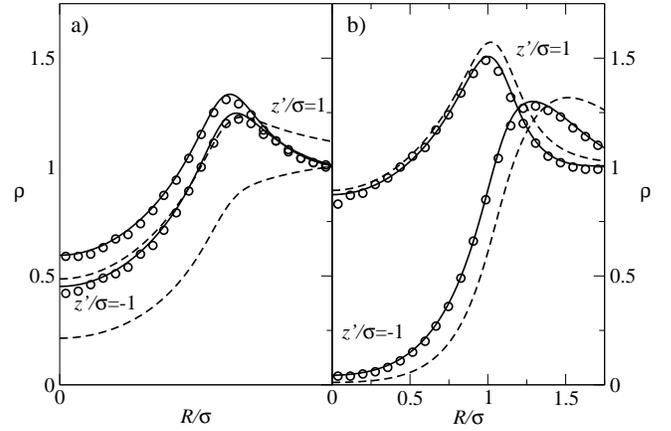}
\caption[]{Steady state density profiles of the polymers around a
  driven colloid at $z'/\sigma=0$ plotted as a function of
  the radial distance $R/\sigma$ from the $z'$-axis for  fixed
  $z'/\sigma=\pm1$. The shifting rates are (a) $\sigma\bar c=1$ and
  (b) $\sigma\bar c=10$. Dashed lines are the results for ideal
  polymers, while the curves for interacting polymers are plotted with
  solid lines. The symbols (circles) are the simulation results.}
\label{fig:2}
\end{figure}
%%%%%%%%%%%%%%%%%%%%%%%%%%%%%%%%%%%%%%%%%%%%%%%%%%%%%%%%%%

Qualitatively we may associate the behavior of the {\it front}  structure to
the direct kinetic effect of the advancing spherical repulsive potential
created by the colloidal particle. As $c$ grows the kinetic constrain
on their brownian trajectories 
becomes the dominant factor for the movement of the polymers at the {\it front} . 
The polymer-polymer interaction plays then a minor role, so 
that for $\bar{c}\sigma \gg 1$ 
the main peak in the {\it front}  structure becomes similar for ideal 
and interacting polymers.  The structure of the  {\it wake}  is determined
by the diffusion from the bulk solution to fill the void left behind the
colloidal particle, the effect of the higher osmotic
pressure accelerates that process and
produces a weaker  {\it wake}  than in the ideal solution limit.
On the opposite extreme, for very low shifting rates of the 
colloidal particle, the effects of the polymer-polymer interactions
are very important, both at the {\it front}  and at the {\it wake}  structures.
At high polymer concentrations the structure around the colloidal particle
is dominated by the steric repulsions between polymers. The
relatively rigid structure of molecular layers is shifted,
with little deformations, by the moving colloidal particle.

\section{Discussion}

As the first point in our discussion, we consider 
the total excess of polymers $\Delta N$, 
produced by $V_{{\rm ext}}({\bf r}')$
over the uniform bulk density.This  is a relevant data since
$c \Delta N$ is the total polymer current,
which requires a total force $ c \Gamma_o \Delta N$
provided by the colloid on the polymers. There is a generic
DDF relation
$$
\Delta N \equiv \int d^3{\bf r}' [\rho({\bf r}')-\rho_o] = 
- \frac{\beta}{\bar{c}} \int d^3{\bf r}' \rho({\bf r}') 
\frac{\partial V_{{\rm ext}}({\bf r}')}{\partial z'},
$$
for the stationary density distributions, which allows to
calculate the total excess from the density distribution
in the neighborhood of the external potential.
The results for the systems in Fig(1) are: $\Delta N_a=3.05$,
$\Delta N_b=1.883$, $\Delta N_c=2.53$ and $\Delta N_d=1.879$. 
Again, the difference between the interacting and the ideal cases
is strongly reduced at $\bar{c}\sigma=10$, with respect to
$\bar{c}\sigma=1$. It is remarkable that the static equilibrium 
result for $\Delta N$ in the ideal gas case would be negative
(as $\rho({\bf r})=\rho_o \exp(-\beta V_{{\rm ext}}({\bf r})) \leq \rho_o)$,
but the stationary excess $\Delta N$ is positive and grows as
$c$ decreases. This may be understood from the analytical 1D result
\cite{1d} for the {\it front}  $\rho(z')-\rho_o\sim  c \, \exp(-\bar{c} z')$, which
 vanishes locally as $c\rightarrow 0$, but it still gives a positive  
integral  over-compensates the depletion inside
the potential barrier and produces $\Delta N>0$, consistently
with the sign (and value) of the total force.  The difference
between the equilibrium ($c=0$) density distribution and that of 
a stationary state at arbitrarily small but positive $\bar{c}$ is 
 remarkable. The  apparent paradox come from 
the concept of stationary state, which would appear
after a short transient period when  $\bar{c}$  is large, but it would
require diverging  times as  $\bar{c}\rightarrow 0$. The very weak
but extended structure of the exponential {\it front}  in  $\rho({\bf r}')$
for $\bar{c} \sigma \ll 1$ would never be observed in practice,
and the transient states $\rho({\bf r},t)$ for any reasonable $t$
would be very similar to the equilibrium structure for $c=0$.
Nevertheless, the strongly anisotropic density distributions, with non-trivial
global effects even for very low $\bar{c}$, suggest important effects on 
the  interaction
between two driven colloids in a bath of quiescent Brownian particles
\cite{joe2}, qualitatively different from effective interactions in
equilibrium \cite{poon,hartmut,likos}.

Finally we comment on the relevance of the  boundary conditions and the 
system dimension by comparing the present results with the $1D$ system 
explored in a previous work \cite{1d}. 
Obviously, in the system explored here
the effects on the bulk polymer solution are limited to the neighbourhood 
of the single colloidal particle. If we consider a finite concentration of 
colloidal particles, all being drifted at the same rate $c$ with respect
to the stationary framework of the solvent, there would be a finite 
induced polymer current per unit volume. The 3D structure, which
offers easy paths for the polymers to escape from the colloidal particles,
would probably  make unfeasible the approach to the full-drift
regime discussed for 1D systems, in which nearly all the particles move
along the shifting potential. Another  possible problem in 3D  
which may be of interest are  currents through a structured barrier
with holes or slits. However, we have to be aware of the intrinsic
limitations  of our DDF approach, particularly in the treatment of the
solvent as 
an inert reference framework for the Langevin dynamics of the polymers,
which is not affected by the shifting external potential 
(or colloidal particle). 
Altogether,  we may conclude that the DDF offers a good theoretical tool
to explore dynamical problems in polymers solutions subject to 
time dependent external potentials which do not affect the 
solvent, and the formalism may also be used to predict, in good agreement 
with BDS, the density structures created around colloidal
particles or similar sized molecules moving slowly with respect to 
the solvent. However, the extension to problems in which the solvent 
plays a more direct role should be regarded with caution, as they may
require a more symmetrical treatment of the solute and the solvent.

This work has been supported by the Direcci\'on General de
Investigaci\'on Cient\'{\i}fica (MCyT) under grant BMF2001-1679-C03-02, 
and a FPU grant AP2001-0074 from the MECD of Spain. JD acknowledges
the financial support of EPSRC within the Portfolio Grant RG37352.

%\newpage


\begin{thebibliography}{10}
\bibitem{poon}
W. C. K. Poon, J.Phys.: Condens. Matter {\bf 14}, R859 (2002).

\bibitem{pusey}
P. N. Pusey, in {\it Liquis, Freezing and the Glass Transition},
ed. by J.-P. Hansen. D. Levesque and J. Zinn-Justin. (North Holland,
Amsterdam, 1991).

\bibitem{AO}
S. Asakura and F. Oosawa, J. Chem. Phys. {\bf 22}, 1255 (1954). 

\bibitem{Ard}
P. G. Bolhius, A. A. Louis, and J.-P. Hansen, Phys. Rev. Lett. {\bf
  89} 128302 (2002).

\bibitem{OurJCP}
U. Marini Bettolo Marconi and P. Tarazona,
J. Chem. Phys. {\bf 110},
8032 (1999).

\bibitem{OurJPCM}
U. Marini Bettolo Marconi and P. Tarazona,
J. Phys. Cond. Matter, {\bf 12},
A413 (2000).

\bibitem{forces}
A. P. Philipse, Curr. Opin. Colloid Interf. Sci. {\bf 2}, 200 (1997);
D. G. Aarts, J. H. van der Wiel, and H. N. W. Lekkerkerker, J. Phys.:
Condens. Matter {\bf 15}, S245, (2003);
P. Wette, et al., J. Chem. Phys. {\bf 114}, 7556 (2001).

\bibitem{1d} F.Penna and P. Tarazona,
J. Chem. Phys. {\bf 119},
1766 (2003).

\bibitem{hartmut}
J.-P. Hansen and H. L{\"o}wen, in {\it Bridging Time Scales: Molecular
  Simulations for the Next Decade}, (Springer, Berlin, 2002), pp. 167-198.

\bibitem{likos}
C. N. Likos, Phys. Rep. {\bf 348}, 267 (2001).

\bibitem{gausspot}
F.H. Stillinger, J. Chem. Phys.{\bf 65}, 3968 (1976).
\bibitem{evanspre}
A.J.Archer and R.Evans, Phys. Rev. E {\bf 64}, 041502 (2001)

\bibitem{evanscm}
A.J.Archer and R.Evans, J.Phys.:Condens. Matter {\bf 14}, 1131 (2002)
\bibitem{lang}
A. Lang, C. N. Likos, M. Watzlawek, and H. L{\"o}wen, 
 J. Phys.: Condens. Matter {\bf 12}, 5087 (2000).

\bibitem{Louis1}
A. A. Louis, P. G. Bolhuis, and J.-P. Hansen, Phys. Rev. E {\bf 62},
7961 (2000)

\bibitem{Louis2}
A. A. Louis, P. G. Bolhuis, J.-P. Hansen, and E. J. Meijer, Phys. Rev. Lett. {\bf 85},
2522 (2000)

\bibitem{joelikos1} J.Dzubiella and C.N.Likos,
J. Phys: Cond. Matter , {\bf 15},
L147 (2003). 

\bibitem{BDS}
M.P Allen  and T.J Tildesley {\it Computer Simulation of Liquids} (Clarendon Press,Oxford 1989) 

\bibitem{joe2}
J. Dzubiella, C. N. Likos, and H. L{\"o}wen, to be published.

\end{thebibliography}
\end{document}